\documentclass{article}
\usepackage{epsfig}

\date{\today}

\normalsize
\begin{document}

\begin{center}
{\Large\bf ON A PETROV-TYPE D HOMOGENEOUS SOLUTION}
\vspace{1.cm}\\
Eugen Radu 
\\ \emph{Department of  Mathematical Physics,
National University of Ireland Maynooth, Ireland}
\\ email: radu@thphys.nuim.ie
\end{center}
\begin{abstract}
We present a new two-parameter family of solutions of Einstein gravity
with negative cosmological constant in 2+1 dimensions.
These solutions are obtained by squashing the anti-de Sitter geometry along one direction
and posses four Killing vectors.
Global properties as well as the four dimensional generalization
are discussed, followed by the investigation of the geodesic motion.
A simple global embedding of these spaces as the intersection of 
four quadratic surfaces in a seven dimensional space is obtained.
We argue also that these geometries describe
the boundary of a four dimensional nutty-bubble solution and
are relevant 
in the context of AdS/CFT correspondence.
\end{abstract}

\section{Introduction}	

Three dimensional (3D) gravity provided us 
with many important clues about higher dimensional physics.
It help that this theory with a negative cosmological constant $\Lambda$ 
has non-trivial solutions, such as the Ba\~nados-Teitelboim-Zanelli
black-hole spacetime \cite{Banados:wn}, which provide important testing ground
for quantum gravity and AdS/CFT correspondence.
Many other types of 3D solutions with a negative cosmological constant 
have also been found by coupling matter fields to gravity
in different way.

The present work was partially motivated by the finding that the nonflat 
part of the famous G\"odel geometry \cite{Godel:1949ga}
can be interpreted as resulting from the squashing of a three dimensional 
anti-de Sitter (AdS) lightcones \cite{Rooman:1998xf}.
Inspired by these results, we consider in this paper the possibility 
of squashing the AdS lightcones along a different direction.
It is possible in this way to find an exact solution 
of 3D Einstein's equations which posseses
the same amount of symmetry as the G\"odel spacetime, without 
the causal pathologies of the later.
This configuration is characterized by two parameters $m$ 
and $n$ and has no curvature singularities.
The metric is given in three different forms, indexed by a parameter $k=0,\pm 1$ which 
are locally equivalent.

We present also a four dimensional (4D) interpretation of this solution; in this case
it satisfies the Einstein-Maxwell-scalar
field equations with a negative cosmological constant.
The 3D solution can be seen as arising from four dimensional  gravity and provided us 
with a clear understanding of how the 3D and 4D gravity are related to each other. 
We have to mention that this spacetime is not asymptotically flat, 
nor AdS. 

The  paper is structured as follows:
in Section 2 we present the derivation of the new line element and 
determine a matter content compatible with this geometry. 
We construct also a global algebraic isometric embedding of these metric 
in a seven dimensional flat space.
In Section 3, a 4D generalization of these solutions is discussed and a matter content 
compatible with Einstein field equations is found. 
The geodesic equations of motion are integrated in Section 4 where 
the properties of trajectories are also discussed.
The Section 5 contains a preliminary discussion of scalar field quantization
for a particular parametrization of this geometry.
The paper closed with Section 6, where our main conclusions and remarks are presented.

We use the same metric and curvature conventions as in \cite{kramer}, and 
we work in units where $c=G=1$.


\section{The geometry and matter content}
\subsection{The line element}
Following Rooman and Spindel \cite{Rooman:1998xf}, we introduce the triad
\begin{eqnarray}
\label{triad}
\theta^1=dx,~~\theta^2=dy+N(x)dt,~~\theta^3=M(x)dt,
\end{eqnarray}
where 
\begin{eqnarray}
\label{MN}
N=\frac{1}{2}(e^{mx}-ke^{-mx}),
~~~M=\frac{1}{2}(e^{mx}+ke^{-mx}),
\end{eqnarray}
and $k=0,\pm 1$. We consider also the set of metrics 
\begin{eqnarray}
\label{metric}
d\sigma^2=
(\theta^1)^2+n^2(\theta^2)^2-(\theta^3)^2.
\end{eqnarray}
Here $n,~m$ are two real parameters; 
for a value  of the squashing parameter $n=1$, 
the above metric describes the geometry of 
$AdS_3$ space (with $\Lambda=-m^2/4$), written in unusual coordinates. 
For example, the transformation
$y=\varphi/m+T/2,~t=\varphi/m-T/2,~x=(2/m)~{\rm arcosh}(mr/2)$
brings the $k=-1$ metric into the more usual form 
\begin{eqnarray}
\label{metric-AdS1}
d\sigma^2=(\frac{m^2r^2}{4}-1)^{-1}dr^2+r^2d\varphi^2-(\frac{m^2r^2}{4}-1)dT^2.
\end{eqnarray}
In this description of (a part of) AdS space, $\varphi$ has to be given the full range, 
$-\infty<\varphi<\infty$ of a hyperbolic angle.
 
By using the rescalling $y \to y/n$ and $N \to nN$ we rewrite (\ref{metric}) as 
\begin{eqnarray}
\label{3D}
d\sigma^2=dx^2+(dy+Ndt)^2-M^2dt^2
\end{eqnarray}
where hereafter $N=n (e^{mx}-ke^{-mx})/2$,
presenting the particular forms
\begin{eqnarray}
\label{k=-1}
d\sigma^2&=&dx^2+(dy+\frac{n}{2}\cosh(mx)dt)^2-\sinh^2(mx)dt^2, ~~{\rm for}\ \ k=-1 
\\
\label{k=0}
d\sigma^2&=&dx^2+(dy+\frac{n}{2}e^{mx}dt)^2-\frac{e^{2mx}}{4}dt^2,~~~~~~~~~~~~~~~~{\rm for}\ \ k=0 
\\
\label{k=1}
d\sigma^2&=&dx^2+(dy+\frac{n}{2}\sinh(mx)dt)^2-\cosh^2(mx)dt^2, ~~~{\rm for}\ \ k=1 .
\end{eqnarray}
At this stage the coordinates $(x,y,t)$ are generic and nothing can be said about the
range of values they take.

As expected, 
line elements with the same value of $m$ and $n$ are isometric
as proven by the existence of the coordinate transformation
\begin{eqnarray}
\label{t1}
\nonumber
\exp{(mx_0)}&=&\exp{(m x_{-1})} \cosh ^{2} (mt_{-1}/2)-\exp{(-m x_{-1})} \sinh ^{2} (mt_{-1}/2), 
\\
t_0\exp{(mx_0)}&=&\frac{1}{m}\sinh (mx_{-1})\sinh (m t_{-1}/2) ,
\\
\nonumber
\tanh \big(\frac{m}{2n}(y_0-y_{-1})\big)&=&\exp{(-m x_{-1})}\tanh(mt_{-1}/2),
\end{eqnarray}
which relates the cases $k=0$ and $k=-1$.
Similarly, a straightforward calculation shows that the transformation 
\begin{eqnarray}
\label{t2}
\nonumber
\exp({mx_0})&=&\cos(mt_1)\cosh(mx_{1})+\cosh(mx_{1}), 
\\
t_0\exp({mx_0})&=&\frac{1}{m}\cosh (mx_{1})\sin (m t_{1}) ,
\\
\nonumber
\tanh \big(\frac{m}{2n}(y_1-y_0)\big)&=&\exp{(-mx_{1})}\tan(m t_{1}/2).
\end{eqnarray}
carries the $k=0 $ metric into $k=1$ one.
In these  relations, the indices of the coordinates
correspond to the value of the parameter $k$.

A standard calculation show that these metrics 
admit at least four Killing vectors. For $k=0$ we find
\begin{eqnarray}
\label{killing0}
\nonumber
K_{1} &=&\frac{\partial}{\partial y},
\\
\nonumber
K_{2} &=&\frac{1}{\sqrt{2}}
\left(\frac{t}{m}\frac{\partial}{\partial x}+\frac{2ne^{-mx}}{m^2}\frac{\partial}{\partial y}
+(1-\frac{1}{2}(t^2+4e^{-2mx}))\frac{\partial}{\partial t}\right),
\\
K_{3} &=&\frac{1}{\sqrt{2}}
\left(\frac{t}{m}\frac{\partial}{\partial x}+\frac{2ne^{-mx}}{m^2}\frac{\partial}{\partial y}
-(1+\frac{1}{2}(t^2+4e^{-2mx}))\frac{\partial}{\partial t}\right),
\\
\nonumber
K_{4} &=&\frac{1}{m}\frac{\partial}{\partial x}-t\frac{\partial}{\partial t}.
\end{eqnarray}
The Killing vectors for $k=1$ metrics are
\begin{eqnarray}
\label{killing1}
\nonumber
K_{1} &=&\frac{\partial}{\partial y},
\\
\nonumber
K_{2} &=&\frac{1}{m}
\left(
\cos(mt)
\frac{\partial}{\partial x}
-\frac{n\sin(mt)}{\cosh(mx)}\frac{\partial}{\partial y}
-\sin(mt)\tanh(mx)\frac{\partial}{\partial t}
\right),
\\
K_{3} &=&\frac{1}{m}\frac{\partial}{\partial t},
\\
\nonumber
K_{4} &=&\frac{1}{m}
\left(
\sin(mt)
\frac{\partial}{\partial x}
+\frac{n\cos(mt)}{\cosh(mx)}\frac{\partial}{\partial y}
-\cos(mt)\tanh(mx)\frac{\partial}{\partial t}
\right),
\end{eqnarray}
while for $k=-1$ we find
\begin{eqnarray}
\label{killing-1}
\nonumber
K_{1} &=&\frac{\partial}{\partial y},
\\
\nonumber
K_{2} &=&\frac{1}{m}\frac{\partial}{\partial t},
\\
K_{3} &=&\frac{1}{m}
\left(
\sinh(mt)
\frac{\partial}{\partial x}
+\frac{n\cosh(mt)}{\sinh(mx)}\frac{\partial}{\partial y}
-\cosh(mt)\coth(mx)\frac{\partial}{\partial t}
\right),
\\
\nonumber
K_{4} &=&\frac{1}{m}
\left(
\cosh(mt)
\frac{\partial}{\partial x}
+\frac{n\sinh(mt)}{\sinh(mx)}\frac{\partial}{\partial y}
-\sinh(mt)\coth(mx)\frac{\partial}{\partial t}
\right).
\end{eqnarray}
These vectors obey the algebra
\begin{eqnarray}
\label{algebra}
[K_1,K_i]=0,~~
\lbrack K_2,K_3 \rbrack=K_4,
~~
\lbrack K_2,K_4\rbrack&=&K_3,
~~
[K_3,K_4]=K_2.
\end{eqnarray}
\subsection{The matter content}
A standard calculation of the Einstein tensor for the metric (\ref{3D})
in the orthonormal triad (\ref{triad}) yields for the nonvanishing components
\begin{eqnarray}
G_x^x=G_t^t=\frac{m^2n^2}{4},~~G_y^y=m^2(1-\frac{3n^2}{4}).
\end{eqnarray}
To find a matter content compatible 
with this geometry,
we couple the Einstein gravity with a negative cosmological constant $\Lambda$,
to an electromagnetic field and a perfect fluid. 
We find that the source free Maxwell equations
\begin{eqnarray}
\label{Maxwell}
\frac{1}{\sqrt{-g}}\frac{\partial}{\partial x^i}(\sqrt{-g}F^{ik})=0 
\end{eqnarray}
admit the simple solution $F^{xt}=c/M$.
However, the corresponding energy momentum tensor 
\begin{eqnarray}
\label{tMaxwell}
T_a^{b(em)}=F_{ac}F^{bc}-\frac{1}{4}\delta_a^bF^2
\end{eqnarray}
takes a simple form for $k=0$ only.
In this case, the expression of the vector potential is $A=A_ydy+A_tdt$, where
\begin{eqnarray}
\label{A}
A_y=cnx,~~A_t=\frac{c(n^2-1)}{2m}e^{mx}.
\end{eqnarray}
The constant $c$ is determined from the Einstein equations 
\begin{eqnarray}
\label{Einstein}
R_a^b-\frac{1}{2}R\delta_a^b+\Lambda_a^b=8\pi T_a^{b}.
\end{eqnarray}
Here the energy-momentum tensor $T_a^b$ 
is the sum of the Maxwell field contribution which in the triad basis (\ref{triad}) reads
\begin{eqnarray}
\label{tem}
T_{x}^{x(em)}=\frac{c^2}{2}(n^2-1),~~T_{y}^{y(em)}=-T_{t}^{t(em)}=\frac{c^2}{2}(n^2+1)
\end{eqnarray}
and the perfect fluid energy-momentum tensor with a general form
\begin{eqnarray}
\label{fluid}
T_a^{b(f)}=(p+\rho)u_au^b+p\delta_a^b,
\end{eqnarray}
where $\rho$ is the energy density of the fluid, $p$ the pressure 
and $u^a$ the three-velocity satisfying $u_a u^a=-1$.
For a pressure-free fluid $(p=0)$ we find
\begin{eqnarray}
\label{c1}
8\pi \rho=-n^2m^2(1-n^2),~~8\pi c^2=m^2(1-n^2),~~
\Lambda=-\frac{m^2}{4}(2n^4-3n^2+2).
\end{eqnarray}
Clearly the condition $n^2<1$ should be satisfied which implies a violation 
of the weak energy condition.
The above solution remains the same 
if we change the sign of $m,~n$. 
As expected, the case $n=1$ corresponds to the $AdS_3$ solution.

Given the existence of the coordinate transformations 
(\ref{t1})-(\ref{t2})
this is a general solution for every value of $k$. However, for 
$k \neq 0$, the expression of the potential vector $A_i$ looks very complicated.

We note also that the cosmological term in Einstein's equations 
can be regarded, if one wishes, as an energy momentum tensor for a perfect fluid.
In this description the cosmological term does not appear explicitly.
The Einstein equations give the relations
\begin{eqnarray}
8\pi \rho=\frac{m^2}{4}(2n^4-n^2-2),~~8\pi c^2=m^2(1-n^2),~~
8\pi p=-\frac{m^2}{4}(2n^4-3n^2+2).
\end{eqnarray}
Since these solutions are not asymptotically flat, nor AdS,
the definition of mass and other conserved quantities is not obvious.
\subsection{Global properties}
The higher dimensional global flat embedding of a curved spacetime 
is a subject of interest to physicists
as well as matematicians.
We mention only that several authors have shown that the global embedding Minkowski approach
of which a hyperboloid in higher dimensional flat space corresponds to
original curved space could provide a unified derivation 
of Hawking temperature for a wide variety of curved spaces (see e.g. \cite{Deser:1997ri}). 

Given the large number of symmetries of the line element (\ref{3D}), 
its geometry takes a simple form
in a large number of coordinate systems, 
which do not usually cover all of the spacetime.

For $n^2<1$ (as requested by the Einstein equations) we found that 
the spacetimes (\ref{3D})
can be introduced using a 7-dimensional formalism as metrics
\begin{eqnarray}
\label{emb1}
ds^2=-(dz^1)^2-(dz^2)^2+(dz^3)^2+(dz^4)^2+(dz^5)^2-(dz^6)^2-(dz^7)^2
\end{eqnarray}
restricted by the constraints
\begin{eqnarray}
\label{s1}
(z^1)^2+(z^2)^2-(z^3)^2-(z^4)^2&=&(\frac{2n}{m})^2,
\\
\label{s2}
\frac{m}{4n^2}\sqrt{1-n^2}\big((z^1)^2-(z^2)^2-(z^3)^2+(z^4)^2\big)&=&z^7,
\\
\label{s3}
\frac{m}{2n^2}\sqrt{1-n^2}(z^1z^4+z^2z^3)&=&z^5,
\\
\label{s4}
\frac{m}{2n^2}\sqrt{1-n^2}(z^1z^2+z^3z^4)&=&z^6.
\end{eqnarray}
The 7-dimensional coordinates $z^i$ are the embedding functions.

Various 3-dimensional parametrizations of these surfaces can be considered.
For example 
\begin{eqnarray}
\nonumber
z^1&=&\frac{2n}{m}\left( \cosh(\frac{mx}{2}) \cosh (\frac{my}{2n})
                               +\frac{1}{2}mte^{mx/2}\sinh(\frac{my}{2n}) \right),
\\
\nonumber
z^2&=&\frac{2n}{m}\left( -\sinh(\frac{mx}{2}) \sinh (\frac{my}{2n})
                               +\frac{1}{2}mte^{mx/2}\cosh(\frac{my}{2n}) \right),
\\
\nonumber
z^3&=&\frac{2n}{m}\left( \cosh(\frac{mx}{2}) \sinh (\frac{my}{2n})
                               +\frac{1}{2}mte^{mx/2}\cosh(\frac{my}{2n} )\right),
\\
z^4&=&\frac{2n}{m}\left( \sinh(\frac{mx}{2}) \cosh (\frac{my}{2n})
                               -\frac{1}{2}mte^{mx/2}\sinh(\frac{my}{2n}) \right),
\\
\nonumber
z^5&=&\sqrt{1-n^2}\frac{1}{4m}\left( 2t^2m^2e^{mx}+4\sinh (mx) \right),
\\
\nonumber
z^6&=&\sqrt{1-n^2}te^{mx},
\\
\nonumber
z^7&=&\sqrt{1-n^2}\frac{1}{4m}\left( -2t^2m^2e^{mx}+4\cosh (mx) \right),
\end{eqnarray}
coresponds to the $k=0$ metric form (\ref{k=0}).
We remark that these coordinates cover entire variety.

Another natural parametrization of (\ref{s1})-(\ref{s4}) is
\begin{eqnarray}
\nonumber
z^1&=&\frac{2n}{m}\cosh(\frac{m x}{2})\cosh\left(\frac{m t }{2}+\frac{m y }{2n}\right),
\\
\nonumber
z^2&=&\frac{2n}{m}\cosh(\frac{m x }{2})\sinh\left(\frac{m t }{2}-\frac{m y }{2n}\right),
\\
\nonumber
z^3&=&\frac{2n}{m}\sinh(\frac{m x }{2})\sinh\left(\frac{m t }{2}+\frac{m y }{2n}\right),
\\
z^4&=&\frac{2n}{m}\sinh(\frac{m x }{2})\cosh\left(\frac{m t }{2}-\frac{m y }{2n}\right),
\\
\nonumber
z^5&=&\frac{\sqrt{1-n^2}}{m}\sinh (m x )\cosh (m t ),
\\
\nonumber
z^6&=&\frac{\sqrt{1-n^2}}{m}\sinh (m x )\sinh (m t ),
\\
\nonumber
z^7&=&\frac{\sqrt{1-n^2}}{m}\cosh (m x ),
\end{eqnarray}
leading to the line-element (\ref{k=-1}).
We remark that in this case the coordinates $(x,~y,~t)$ covers only a half of the hyperboloid 
(\ref{s1}) since $z^1>z^2,~z^3>z^4$.
Thus, these coordinates are an analogous to Rindler 
coordinates of flat space and need to be analytically extended
in the usual fashion to cover all of the spacetime.

The parametrization corresponding to a metric form  with $k=1$ is
\begin{eqnarray}
\nonumber
z^1&=&\frac{2n}{m}\left(
\cosh(\frac{m x}{2})\cosh(\frac{m y}{2n})\cos(\frac{m t}{2})+
\sinh(\frac{m x}{2})\sinh(\frac{m y}{2n})\sin(\frac{m t}{2}) \right),
\\
\nonumber
z^2&=&\frac{2n}{m}\left(
\cosh(\frac{m x}{2})\cosh(\frac{m y}{2n})\sin(\frac{m t}{2})-
\sinh(\frac{m x}{2})\sinh(\frac{m y}{2n})\cos(\frac{m t}{2}) \right),
\\
\nonumber
z^3&=&\frac{2n}{m}\left(
\cosh(\frac{m x}{2})\sinh(\frac{m y}{2n})\cos(\frac{m t}{2})+
\sinh(\frac{m x}{2})\cosh(\frac{m y}{2n})\sin(\frac{m t}{2}) \right),
\\
\nonumber
z^4&=&\frac{2n}{m}\left(
-\cosh(\frac{m x}{2})\sinh(\frac{m y}{2n})\sin(\frac{m t}{2})+
\sinh(\frac{m x}{2})\cosh(\frac{m y}{2n})\cos(\frac{m t}{2}) \right),
\\
z^5&=&\frac{\sqrt{1-n^2}}{m}\sinh (m x),
\\
\nonumber
z^6&=&\frac{\sqrt{1-n^2}}{m}\cosh (m x)\sin(m t),
\\
\nonumber
z^7&=&\frac{\sqrt{1-n^2}}{m}\cosh (m x)\cos(m t).
\end{eqnarray}
We mention also, without entering into details, the  parametrization
\begin{eqnarray}
\nonumber
z^1&=&\frac{2n}{m}\cos(\frac{mx^4}{2})\cosh(\frac{m(x^1+x^2)}{2n}),
\\
\nonumber
z^2&=&\frac{2n}{m}\sin(\frac{mx^4}{2})\cosh(\frac{m(x^1-x^2)}{2n}),
\\
\nonumber
z^3&=&\frac{2n}{m}\cos(\frac{mx^4}{2})\sinh(\frac{m(x^1+x^2)}{2n}),
\\
z^4&=&\frac{2n}{m}\sin(\frac{mx^4}{2})\sinh(\frac{m(x^1-x^2)}{2n}),
\\
\nonumber
z^5&=&\frac{\sqrt{1-n^2}}{m}\sin(mx^4)\sinh(\frac{mx^1}{n}),
\\
\nonumber
z^6&=&\frac{\sqrt{1-n^2}}{m}\sin(mx^4)\cosh(\frac{mx^1}{n}),
\\
\nonumber
z^7&=&\frac{\sqrt{1-n^2}}{m}\cos(mx^4),
\end{eqnarray}
which gives a time-dependent line element
\begin{eqnarray}
\label{par3}
d\sigma^2=\frac{\sin^2 (mx^4)}{n^2} ~(dx^1)^2+(dx^2+\cos( mx^4)~dx^1)^2-(dx^4)^2
\end{eqnarray}
The transformation rules between the various coordinates may be easily obtained
by comparing their definitions
in terms of the basic embedding
coordinates $z_i$.

Note that the embedding presented in this section can easily be extended for $n^2>1$ or $m^2<0$.
%
\section{The solution in D=4}
\subsection{The Einstein equations}
The metric (\ref{3D}) can be added to a $(D-3)$ dimensional Euclidean
 metric $d\Sigma^2_{D-3}$ to give a $D-$dimensional generalization 
of this geometry. 
This can be achieved by expressing the $D-$dimensional metric as the direct Riemannian sum 
\begin{equation} 
\label{D}
ds^2=d\sigma^2+d\Sigma^2_{D-3}.
\end{equation}
The physical interesting case $D=4$ has a particularly simple matter content.
The corresponding line elements read
\begin{equation} 
\label{D4}
ds^2=dx^2+dy^2+dz^2+2N(x)dydt-(M(x)^2-N(x)^2)dt^2.
\end{equation}
Here $z$ is a Killing direction in 4D spacetime and has the natural range $-\infty<z<\infty$.
In this case the computations are done in a orthonormal tetrad basis $\omega^a=(\theta^a,dz)$.
In this basis, the only new components of the Einstein tensor is $G_z^z=m^2-m^2n^2/4$.
The determinants of the metrics are the same $g^{(4)}=g^{(3)}$ and 
also the Ricci scalars.

We find that (\ref{A}) still solves the Maxwell equations.
However, in order to satisfy the $(zz)$ Einstein equation we 
have to include a massless scalar field in the theory,
apart from a cosmological constant and a perfect fluid.
A particular solution of the field equation for a massless minimally 
coupled scalar field 
\begin{eqnarray} 
\frac{1}{\sqrt{g} } \frac{\partial }{\partial x^{i} } (\sqrt{g} g^{ij}
\frac{\partial \Psi }{\partial x^{j} } )=0
\end{eqnarray} 
is $\Psi =ez$, which implies the nonvanishing components of energy-momentum tensor
\begin{eqnarray} 
T_x^{x(s)}=T_y^{y(s)}=-T_z^{z(s)}=T_t^{t(s)}=-\frac{e^2}{2}.
\end{eqnarray} 
For $k=0$ and a pressure free perfect fluid, we find from the 4D Einstein equations 
\begin{eqnarray} 
\nonumber
8\pi c^2&=&m^2(1-n^2),~8\pi e^2=\frac{m^2n^2}{2}(3-2n^2),
\\
8\pi \rho&=&-m^2 n^2(1-n^2),~~\Lambda=-\frac{m^2}{2}.
\end{eqnarray} 
We remark that the matter content (\ref{c1}) of the 3D solution is obtained 
through a compactification along the $z-$direction
of the 4D solution. 
The only effect of the scalar field is to shift the value of the cosmological constant 
$\Lambda^{(3)}=\Lambda^{(4)}+4\pi e^2$.

However, in view of the fact that the cosmological term can be regarded as 
a perfect fluid, we can again switch the 4D
description into a fluid with the energy density given by $8\pi \rho=-m^2(3-2n^2)/2$ and a 
nonzero pressure $8\pi p=m^2/2$.
The values of $c$ and $e$ obtained above are still valid. 
Again, the matter content for $k=\pm 1$ can be found by using 
the covariance of the field equations.

In four dimensions we find also a different solution of the Maxwell equations, 
corresponding to a vector potential
\begin{eqnarray} 
A=-\frac{E}{mn} (dy+Ndt),
\end{eqnarray}
where $E$ has a constant value.
Different than (\ref{A}), this gives a simple matter content compatible 
with the geometry (\ref{metric}) for every value of $k$.
Also, we use the same solution of the Klein-Gordon equation, $\Psi=ez$.
In this approach, the matter content is given by a Maxwell and a scalar field  
($i.e.$ no perfect fluid),
the Einstein equations with cosmological constant
implying to the following relations
\begin{eqnarray} 
\label{matter}
4\pi E_0^2=\frac{m^2}{2}(1-n^2),~~
4\pi e^2 =\frac{m^2n^2}{4},~~
\Lambda = -\frac{m^2}{2}.
\end{eqnarray}
Clearly the relation $n^2<1$ should again be satisfied. 
However, this time all three energy conditions are satisfied.

To examine the Petrov classification of the four line element 
(\ref{D4}) a complex 
null tetrad was chosen having the tetrad basis ($i=\sqrt{-1}$)
\begin{eqnarray}
\nonumber
\omega ^{1} &=&\bar{m} _{i} dx^{i} =\frac{1}{\sqrt{2} } (dx-idz),
\\
\nonumber
\omega ^{2} &=&m_{i} dx^{i} =\frac{1}{\sqrt{2} } (dx+idz), 
\\
\omega ^{3} &=&\frac{1}{\sqrt{2} } \Big((M(x)-N(x)) dt-dy\Big),
\\
\nonumber
\omega ^{4} &=&\frac{1}{\sqrt{2} } \Big((M(x)+N(x)) dt+dy\Big). 
\end{eqnarray}
In this null tetrad, the Weyl 
tensor invariants \cite{kramer} are 
$
\Psi _{0} =-3\Psi _{2}=\Psi _{4}=m^2(1-n^{2})/4,~~
\Psi _{1} =\Psi _{3}=0. 
$
Therefore the metric is Petrov type $D$, except the case $n=1$ 
which is Petrov type $N$. 
\subsection{Limiting cases and relation with known solutions}
The diagonal limit of this solution is obtained for $n=0$
\begin{eqnarray} 
ds^2=dx^2+dy^2+dz^2-M(x)^2dt^2.
\end{eqnarray}
For $k=\pm1$ this corresponds to scalar worlds discussed in \cite{dariescuk=1} 
(the case $M=\cosh(mx)$) and 
\cite{dariescuk=-1} (the case $M=\sinh(mx)$), where the geodesic equations are solved
and the pathological features of these solutions are examined.
The influence of the global properties on the behaviour of magnetostatic fields in such universes
is also studied.
We mention the existence of one more symmetry in this case, corresponding to
a rotation in $yz$ plane.

The four dimensional solution with $k=1,~n=1$ is well known in the literature. 
It corresponds to
Rebou\c{c}as-Tiomno space-time, originally found as a first example of 
G\"odel-type homogeneous solution without CTCs \cite{reboucas}.
The properties of this line element are discussed in \cite{Reboucas:wa}, 
where the geodesic equations
are integrated. The solution in this case possesses seven isometries and is conformally flat.


In fact, the four dimensional line element (\ref{D4}),
the two-parameter G\"odel-type homogeneous solution \cite{reboucas} and the static Taub
solution  \cite{Taub:1951ez} corresponds to different analytical continuations
of the same euclidean line element
\begin{eqnarray}
\label{euclid}
ds^{2}_E =dx^2+(dy+\frac{\bar{n}}{2}(e^{mx}-ke^{-mx})d\tau)^2+dz^2
+\frac{1}{4}(e^{mx}+ke^{-mx})d\tau^2.
\end{eqnarray}
where again $k=0,\pm 1$. 
The global properties of this line element can easily be obtained following 
the results presented in Section (2.3).

The general line element (\ref{D4}) is obtained by anlytical continuation
$\tau \to it,~\bar{n} \to in$.  
The analytical continuation $\tau \to i\varphi/m,~y \to i(T-n\varphi/m)$ 
of the $k=-1$ line element (\ref{euclid}) yields
the most general expression of a homogeneous G\"odel-type space-time
in cylindrical coordinates
\begin{equation} 
\label{godel}
ds^2=dr^2+\frac{\sinh^2 m r}{m^2}d\varphi^2+dz^2-\Big(\frac{4\Omega}{m^2}
\sinh^2(\frac{m r}{2})d\varphi+dT\Big)^2,
\end{equation}
after the identification $x=r,~n=2\Omega/m$. 
A symilar analytical continuation of the $k=0$ line element (\ref{euclid}) yields a 
G\"odel-type solution in cartezian coordinates.
The properties of a G\"odel-type space-time were discussed by many authors 
(see e.g. \cite{Obukhov:2000jf} for a large list of references). 
However, the properties of the solution proposed 
in this paper differs form those of an homogeneous rotating G\"odel-type space-time, 
excepting the common case $n^2=1$.

Another case which we should 
mention is the static Taub line element \cite{Taub:1951ez}, corresponding to the analytical 
continuation $z \to iT$ of the 
line element (\ref{euclid}).
\subsection{Causal properties} 
One of the features of the AdS spaces is that they admit closed timelike curves (CTC). 
The usual remedy for this is to consider
the covering space instead of AdS itself.
Looking at (\ref{s1}) we see that our geometry suffers from the same problem and it admits CTC
(remember that $z^1$ and $z^2$ both are timelike coordinates).
However, these CTC are already present 
in $AdS_3$ spacetime and can be removed by considering the covering space.

Apart from this property, this solution is geodesically complete and satisfy 
causality conditions such as global hyperbolicity. 
Different from the G\"odel solution, a cosmic time function can be defined;
for the function $f=t$, one has
\begin{eqnarray}
g^{ij} \frac{\partial f}{\partial x^{i} } \frac{\partial f}{\partial
x^{j} } =-\frac{1}{M(x)^{2}} <0
\end{eqnarray}
for every finite value of $x$,
implying that CTC are not present.

Since in $D=4$ a perfect fluid is not necessarily present as source of curvature, the 
kinematical parameters of the model are not unambiguously defined. 
However, we can consider  an observer with a four-velocity given by 
$
u^{i} u_{i} =-1
$ with
$
u^{i} =\delta_{4}^{i}(M^2-N^2)^{-1/2}
$
and find  no expansion ($\theta=0$), no shear ($\sigma_{ij}=0$) but a non-null 
vorticity
\begin{eqnarray}
\omega ^{i} =\frac{1}{2\sqrt{-g} } k ^{ijkm} u_{j;k} u_{m}
=\frac{1}{2(M^2-N^2)}\left(2NM'-mn(M^2+N^2)\right)\delta_{z}^{i}
\end{eqnarray}
For the case $k=0$ we find a vorticity parallel to the $z$ axis
of magnitude $mn/2$.

From the expressions above it is obvious that the spacetime
(\ref{D4}) has no curvature singularities anywhere and also its 3D version.
The coordinate ranges can be taken $-\infty<x,y,z,t<\infty$ except for $k=-1$.

While the properties of the $k=0$ and $k=1$ line elements are very similar,
the case $k=-1$ is rather special.
For $k=-1$ the surface $x=0$ presents all the features of an event horizon.
In the limit of no matter $(n=0)$ and zero cosmological constant $(m=0)$ 
we obtain the Rindler spacetime after a rescaling $t \to t/m$.
All the properties of the Rindler solution are shared by this spacetime.
As a new feature, we remark the occurence  of an ergoregion, induced by
the presence of a squashing parameter $n<1$.
We can see from (\ref{k=-1}) that in this case, 
the Killing vector $\partial/\partial t$ is spacelike for $\tanh^2(mx)<n^2$.
These properties are manifest when integrating the geodesic equation.

\section{Geodesic motion and properties of the trajectories}
The study of timelike and null geodesics is an adequate
way to visualize the main features of a spacetime.
In this section we want to study the geodesic motion for $n^2<1$ and, 
in particular, to confirm that the spacetime
described by (\ref{D4}) is both null and timelike geodesically complete.
The metric symmetries  enable us to solve directly some
the motion equations for timelike and null geodesics for every $k$.

For the general line element (\ref{D4}),
the equations of 
the geodesics have the four straightforward first integrals
\begin{eqnarray}
\nonumber
P_{y}&=&\dot{x} _{2} =\dot{y} +N\dot{t} , 
\\
\nonumber
P_{z}&=&\dot{x} _{3} =\dot{z}, 
\\
\label{int3}
E&=&\dot{x} _{4} =N\dot{y}+(N^2-M^2)\dot{t},
\\
\nonumber
-\varepsilon&=&\dot{x} ^{2} +\dot{z} ^{2} +(\dot{y}+N\dot{t})^{2} 
-M^2(x)\dot{t}^{2} 
\end{eqnarray}
where a superposed dot stands for as derivative with respect 
to the parameter $\tau$ and $\varepsilon=1$ or 
$0$ for timelike or null geodesics respectively. 
$\tau$ is an affine parameter along the geodesics; for timelike geodesics, 
$\tau$ is the proper time.
The corresponding relations for $D=3$ are obtained by setting $Py=0$.
The first three 
integrals in (\ref{int3}) are due to the existence of the Killing vector 
fields $\partial _{y},\partial _{t},\partial _{z} $ respectively.
The fourth one is related to the 
invariance of the timelike or null character of a given geodesic, while
$P_y,~P_z$ and $E$ are the corresponding constants of motion.
Note that since the solution is not asymptotically flat, nor AdS, these constants
cannot be identified as the linear momentum and the energy at infinity.

From the above equations we get the simple relations
\begin{eqnarray}
\label{g1}
M^2\dot{x}^2=(NP_y-E)^2-M^2(\varepsilon +P_y^2+P_z^2)
\end{eqnarray}
and
\begin{eqnarray}
z=P_{z} (\tau -\tau_{0} )
\end{eqnarray}                                                            
To integrate the eq. (\ref{g1}) it is convenient to introduce the variable 
$u=NA+P_yE/A$, where
\begin{eqnarray}
A=\frac{\sqrt{\varepsilon +P_y^2(1-n^2)+P_z^2}}{n}.
\end{eqnarray}   
This yields the parametric form of the $x-$coordinate
\begin{eqnarray}
\label{eq-x}
N(x) = a+b\sin \alpha(\tau-\tau_0)
\end{eqnarray}
where
\begin{eqnarray}
\alpha=mnA,~~a=-\frac{EP_y}{A^2},~~
b=\frac{\sqrt{(\varepsilon+P_y^2+P_z^2)(\frac{E^2}{n^2A^2}-k)}}{A}.
\end{eqnarray}
Therefore, for $k=1$, the geodesic motion is possible for 
$E^2>\varepsilon+P_y^2(1-n^2)+P_z^2$ only.
Equation (\ref{eq-x}) shows that, for any value of the constants of motion, 
the massive particles and photons are always confined in a finite region along 
the $x$ axis.

We can easily find a closed form relation between the coordinates $x$ and $t$. Thus
\begin{eqnarray}
\label{x-t}
\nonumber
(\frac{NP_y-E}{M})^2&=&m^2E^2(t-t_0)^2+\varepsilon+P_y^2+P_z^2,
~~~~~~~~~~~~~~~~~~~~~~~~~~{\rm for}\ \ k=0
\\
\frac{EN+n^2P_y}{M}&=&n\sqrt{E^2-(\varepsilon+P_y^2(1-n^2)+P_z^2)}\sin(m(t-t_0)),~~~~{\rm for}\ \ k=1
\\
\nonumber
\frac{EN-n^2P_y}{M}&=&n\sqrt{E^2+\varepsilon+P_y^2(1-n^2)+P_z^2}\sinh(m(t-t_0)) ~~~~~~{\rm for}\ \ k=-1.
\end{eqnarray}
For $k=0,-1$, the equation for $y(\tau)$ gives
\begin{eqnarray}
\label{y-tau}
y-y_0=P_y(1-n^2)(\tau-\tau_0)
-\frac{n^3P_y}{2\alpha}(I_{+}(k)-I_{-}(k))
+\frac{n^2E}{2\alpha}(I_{+}(k)+I_{-}(k))
\end{eqnarray}
where
\begin{eqnarray}
\nonumber
I_{\pm}(k)=
\frac{1}{\sqrt{b^2-(a\pm kn)^2}}
\ln\left(
            {
\frac{
(a\pm kn)\tan\big(
\alpha(\tau-\tau_0)
               /2\big)
+b-\sqrt{b^2-(a\pm kn)^2}
     }
     {(a\pm kn)\tan\big(
\alpha(\tau-\tau_0)
                  /2\big)
+b+\sqrt{b^2-(a\pm kn)^2}}}
\right),
\end{eqnarray}
while for $t(\tau)$ we find
\begin{eqnarray}
t-t_0&=&\frac{n(nP_y+E)}{2}I_{-}(-1)+\frac{n(nP_y-E)}{2}I_{+}(-1),~~~~{\rm for}\ \ k=-1
\\
\nonumber
t-t_0&=&\frac{2b\alpha}{nm^2E}\frac{\cos(\alpha(\tau-\tau_0))}{a+b\sin(\alpha(\tau-\tau_0))}~~~~{\rm for}\ \ k=0.
\end{eqnarray}
The expresion of $y(\tau)$ and $t(\tau)$ for  $k=1$ 
can also be obtained from (\ref{int3}) but are very 
complicated and we prefer do not present them here since the conclusions in this case
are rather similar to the case $k=0$.
The equation for $y$ can also be read from the folllowing straightforward integral
\begin{eqnarray}
\label{g11}
\int\dot{x}^2 d \tau+(y-y_0)P_y+(z-z_0)P_z+(t-t_0)E= -\varepsilon (\tau-\tau_0).
\end{eqnarray}
It is obvious from the above relations  that for $k=-1$, the surface
$x=0$ presents all the 
characteristics of an event horizon. 
For a freely falling observer an infinite time $t$ is required to traverse 
the finite distance $L_0$ between an exterior point and a point on the horizon, 
but that destination is reached in a finite proper time. 
The running backwards of $t$ for some intervals of $\tau$ has nothing 
to do with a possible going
backward in time or time travel.
This effect is a mere consequence of the special choice of the time coordinate.

\section{Quantum effects} 
The line element (\ref{3D}) has also another interesting property, 
being connected to the a special class of bubble spacetimes \footnote{A 
general clasification of bubbles in (anti-) de Sitter spacetime
can be found in \cite{Astefanesei:2005eq}.}.
The four dimensional "topologically nutty bubbles" obtained
by Ghezelbash and Mann in \cite{Ghezelbash:2002xt}
as a suitable analytical continuation of a Taub-Nut-AdS geometry
can be written in a compact way as
\begin{equation} 
\label{nut}
ds^2=\frac{dr^2}{F(r)}+F(\chi)\left(d\chi+2\tilde{n}\frac{df_k(\theta)}{d \theta} dt\right)^2
+(r^2+\tilde{n}^2)(d\theta^2-f_k^2(\theta)dt^2),
\end{equation}
where 
\begin{eqnarray}
F(r)=\frac{r^4+(-\ell^2+6\tilde{n}^2)r^2-2m \ell^2r-\tilde{n}^2(-\ell^2+3\tilde{n}^2)}
{\ell^2(r^2+\tilde{n}^2)}.
\end{eqnarray}
The discrete parameter $k$ takes the values $1, 0$ and $-1$ 
and implies the form of the function $f_k(\theta)$
\begin{equation}
f_k(\theta)=\frac{1}{2}(e^{\theta}+ke^{-\theta}).
\end{equation}
Here $m$ is the mass parameter, $r$ a radial coordinate, 
$\tilde{n}$ the nut charge and $\Lambda=-3/\ell^2$ the cosmological constant.
The $\theta$ coordinate is no longer periodic and takes on all real values.
One can easily see that a hypersurface of constant large 
radius $r$ in this four-dimensional 
asymptotically AdS spacetime
has a metric which is proportional to the three-dimensional line element (\ref{3D})
after the identifications $x=\theta l$; $m=1/\ell$; $n=2\tilde{n}/\ell$.

The Maldacena conjecture \cite{Maldacena:1998re} 
implies that a theory of quantum gravity 
 in a $(D+1)$-dimensional spacetime with a negative cosmological constant
can be modeled by the a conformal
field theory in the fixed $D$-dimensional boundary geometry. 
Therefore the interest in field quantization in the background (\ref{3D}),
since it will encode the informations on the quantum gravity in a "topologically nutty bubble".
Although the corresponding field theory
 is not known in this case, similar to other situations one may consider
the simple case of a nonminimally coupled scalar field.

One approach toward field quantization is to work directly in the (Lorentz signature)
spacetime under consideration. This approach has the advantage
of yielding a direct, physical interpretation of the results obtained.
The line element (\ref{D4}) presents a global $t=const.$ Cauchy 
surface and the standard methods of quantization 
can be directly applied \cite{Birrell:ix}.
Due the high degree of symmetry it is 
possible to solve the scalar wave equation in terms of hypergeometric functions
for any value of $k$.

An alternative approach  is to
define all quantities on a Euclidean manifold 
(i.e. a positive definite metric). 
The results on the Lorentzian section are to be obtained by analytical 
continuation of the Euclidean quantities.
Here we remark that the "nutty bubble" geometry (\ref{nut}) 
and the general $D=4$ Taub-NUT-AdS family of solutions
discussed in \cite{Astefanesei:2004kn}  
share the same Euclidean section.
The Euclidean boundary geometry in both cases corresponds to the line
element (\ref{euclid}).
Therefore a number of general results 
found in \cite{Astefanesei:2004kn} working on the Euclidean section
are valid in this case too (in particular the computation of the solutions' action).

Concerninig the field quantization, the results for
(the essential three-dimensional part of-) a G\"odel universe and 
 the new 3D solution will correspond
to different analytic continuation of the same Euclidean quantities
(a similar correspondence exists 
$e.g.$ between  the quantization in
Rindler spacetime and a cosmic string background, see e.g. \cite{Moretti:1997qn}).
The Euclidean approach would enable us to use the 
powerful formalism of 
``the direct local $\zeta$-function approach'' \cite{Moretti:1997qn}, and 
to compute the effective action, the vacuum fluctuation 
and the one-loop renormalized stress tensor for a quantum field 
propagating in the background (\ref{3D}).
We remark that
for $k=1$, the general $\zeta$-computation presents some similarities with the
squashed three-sphere case discussed in \cite{Dowker:1998pi}, the case $k=0$
being approached in Appendix B of Ref. \cite{Astefanesei:2004kn}.
A $\zeta$-computation of the effective action for a 
conformal scalar field propagating in the line element (\ref{3D}) 
will be presented elsewhere, since a 
separate analysis is required for every $k$. 

For the rest of this Section, we present 
instead some preliminary results concerning the case 
$k=0$ in four spacetime dimensions,
in which case the results have a particularly simple form,
 focussing on the Green's function computation for a massive scalar field.


A 4D Euclidean line-element is obtained from (\ref{k=0}) by using the 
analytical continuation $t\to -i\tau$ and 
 $n \rightarrow i\bar{n } $
\begin{eqnarray}
ds^2=dx^2+(dy+\frac{\bar{n}}{2}e^{mx}d\tau)^2+dz^2+\frac{1}{4}e^{2mx}d\tau^2.
\end{eqnarray}
The Feynman propagator for a massive scalar field is found
by taking the (unique) solution bounded on the Euclidean section 
of the inhomogeneous equation
\begin{eqnarray}
(\nabla _{a} \nabla ^{a} -M^{2})G_{E} (x,x')=-\frac{\delta
^{4} (x,x')}{g^{1/2} (x)}, 
\end{eqnarray}
the case of a nonminimally coupled scalar field corresponding to a 
particular value of $M^2$.
If the field is at zero temperature $G_E(x, x')$ has the form
\begin{eqnarray}
G_{E} ( x,x')=\frac{1}{8\pi ^{3} } \int\limits_{-\infty }^{+\infty
}d\omega \int\limits_{-\infty }^{+\infty }dk_{y} \int\limits_{-\infty
}^{+\infty }dk_{z}   e^{-i\omega (\tau -\tau' )}  e^{ik_{y} (y-y')}
e^{ik_{z} (z-z')} f_{k_{y} k_{z} \omega } (x,x'),
\end{eqnarray}
and the only remaining equation for the propagator is
\begin{eqnarray}
\label{eq-f}
e^{-mx}\frac{d}{dx}\left(e^{mx}\frac{df}{dx}\right)-4\omega^2e^{-2mx}f-4\bar{n}e^{-mx}k_y\omega f
\\
\nonumber
-\left((1+\bar{n}^2)k_y^2+k_z^2+M^2\right)f\sim -\delta.
\end{eqnarray}
By using the substitution 
$
u=4|\omega| e^{-mx}/m,
$
 the solutions to eq. (\ref{eq-f}) when 
the right hand is zero are
\begin{eqnarray}
\nonumber
f_{1} =M_{k\mu } (u)=e^{-\frac{u}{2} } u^{\mu +\frac{1}{2} }~_1F_{1} \left(\mu
-k+\frac{1}{2},1+2\mu ,u\right),
\end{eqnarray}
finite as $u \to 0$ and
\begin{eqnarray}
\nonumber
f_{2} =W_{k\mu }(u)=e^{-\frac{u}{2} } u^{\mu +\frac{1}{2} } U\left(\mu
-k+\frac{1}{2} ,1+2\mu ,u\right), 
\end{eqnarray}
finite as $u \to \infty$.
In the relations above
$\mu =\sqrt{1/4 +\left(k_{z}^{2}+
(1+\bar{n}^2)k_y^2+M^2\right)/m^2}$, 
$k =-|\omega|\bar{n}k_y/\omega m$, 
$_1F_1$ and $U$ are confluent hypergeometric functions.
The Whittacker functions satisfy the relation \cite{grad}
\begin{eqnarray}
\label{transf}
W_{k\mu } (z)=\frac{\Gamma (-2\mu )}{\Gamma (\frac{1}{2} -\mu -k)}
M_{k\mu } (z)+\frac{\Gamma (2\mu )}{\Gamma (\mu +\frac{1}{2} -k)} M_{k-\mu
} (z).
\end{eqnarray}
Thus the expressions for the spatial part of the Green's function 
is given by the usual expression \cite{morse}
\begin{eqnarray}
f(u,u')=\frac{f_{>} (u_{>} )f_{<} (u_{<} )}{W(f_{>} ,f_{<} )} ,
\end{eqnarray}
where 
 $f_{>}=f_{2}$ satisfies the boundary condition of finiteness at large 
$u$ and $f_{<}=f_{1}$ is similar finite as $u$ goes to zero; 
$W$ is the wronskian 
of $f_>$ and $f_<$ . Any other combination of the linearly independent 
solutions will not satisfy the boundary conditions. However, 
it is necessary to check that no Euclidean bound states exist. 
If an everywhere-finite solution of the homogeneous equation 
do exist, then the freedom one has in adding an arbitrary solution 
of the homogeneous equation (satisfying the boundary conditions) 
to the Green's function would make the Green's function nonunique. 
The condition of existence for a bounded state is 
\begin{eqnarray}
k-\mu-\frac{1}{2}=\rm{positive~integer}.
\end{eqnarray} 
It is easy to see that this cannot happen in the situation under 
consideration.

Returning to the Lorentzian section by continuing back from the 
Euclidean values to the Lorentzian values we can define a Feynman 
propagator. The choice of the above analytical continuation fixes 
the appropriate sign of the timelike Klein-Gordon norms of $f_1$ 
and $f_2$.

The $G_E(x, x')$ contains all the information about the theory \cite{Birrell:ix}. 
As a simple application, we use
the formalism proposed by Lapedes in \cite{Lapedes:1978rw} to prove
 that an inertial observer at constant $x$ will not see any 
particles, where ``what an observer sees`` means ``how a detector 
reacts when it is coupled linearly to a quantized field propagating 
freely in our space-time``. 
In Ref.\cite{Lapedes:1978rw} it has been proved that the average number of produced 
pairs detected by an observer at constant $x$ is
\begin{eqnarray}
<n_{\omega k_{y} k_{z} } >=\frac{w}{1-w},                                                            
\end{eqnarray}
where $w$ is the probability for one pair to be created in a state 
characterized by the quantum numbers $k_y,~ k_z,~\omega$. This probability 
can be computed by returning to the Lorentzian section and writing
\begin{eqnarray}
f_{<} =A\bar{f}_{>} +Bf_{>} .
\end{eqnarray}
The coefficients $A$ and $B$ can be read from (\ref{transf}). Thus 
 $w=1-\frac{\left| B\right| ^{2} }{\left| A\right| ^{2} } $ 
and, for situation discussed here with $n^2<1$, the relative probability is null.

\section{Conclusions}
This paper was inspired by the finding that the G\"odel geometry
can be obtained by squashing the three dimensional 
anti-de Sitter geometry.
Although the initial metric form to be squashed cannot be entirely arbitrarly, 
the variety of possibilities and of the resulting spacetimes is quite large.
 
In this way we obtained a new family of solutions of the 3D Einstein's 
field equations with negative cosmological constant.
This solution is characterized by two continuos parameters $m$ 
and $n$ and a discrete parameter $k$.
So far, physically acceptable sources for these solutions are found 
only for $n^2<1$. 
In a four dimensional interpretation
it satisfies the Einstein-Maxwell-scalar
field equations with cosmological constant.
This space-time, of course, is not a live candidate for describing 
a physical situation, but it can be a source of insight into 
the possibilities allowed by relativity theory. 

We have presented a global description of the spacetime geometries of our solution
by isometrically embedding it in a flat spacetime with four extra dimensions.
In this way we gained a rather clear global structure of the geometry.
A detailed study of the geodesics of this spacetime showed 
that the solution is geodesically complete
and therefore singularity free.

Although this new family posseses
the same amount of symmetry as an homeogeneous G\"odel-type spacetime
(in fact we prove it corresponds to a suitable analitical continuation of the latter),
there are some important differences.
The most obvious difference is the absence of CTC
and a different geodesic structure.
We noticed a relevance of this 3D geometry within AdS/CFT correspondence since it is
the boundary of the four dimensional "topologically nutty bubbles" with negative cosmological 
constants discussed in \cite{Ghezelbash:2002xt}.

Since the solution presented here contains three parameters which may lead to spacetimes
with rather different properties, it would be interesting to extend the analysis 
for $n^2>1$ and $-\infty<m^2<\infty$.
Another interesting problem is to study the properties of solutions obtined from 
other parametrizations
of the quadratic surfaces (\ref{s1})-(\ref{s4}).
\\
\\
{\bf Acknowledgement}
\newline
The author is grateful to C. Dariescu for useful remarks.
 This work was performed in the framework of Enterprise--Ireland Basic Science Research
Project
SC/2003/390 of Enterprise-Ireland.


\end{document}